\documentclass[11pt]{article}

\usepackage{cite}             
\usepackage{amssymb}          
\usepackage{latexsym}         
\usepackage{amsmath}          
\usepackage[latin1]{inputenc} 
\usepackage{eucal}            
\usepackage{bbm}              
\usepackage{theorem}          
\usepackage[ps,dvips]{xypic}  
\usepackage{diagxy}           

\title{The Covariant Picard Groupoid in Differential
  Geometry\thanks{Talk given at the 20th International Workshop on
    Differential Geometric Methods in Theoretical Mechanics in Ghent,
    August 2005.}  }

\author{\addtocounter{footnote}{5} \textbf{Stefan
    Waldmann}\thanks{E-mail: Stefan.Waldmann@physik.uni-freiburg.de}
  \\[0.1cm]
  Fakult{\"a}t f{\"u}r Mathematik und Physik\\
  Albert-Ludwigs-Universit{\"a}t Freiburg\\
  Physikalisches Institut\\
  Hermann Herder Stra{\ss}e 3\\
  D 79104 Freiburg\\
  Germany}

\date{September 2005\\[0.5cm] FR-THEP 2005/10}

\renewcommand{\mathbb}[1]{\mathbbm{#1}} 

\newcommand{\id}         {\operatorname{\mathsf{id}}}   
\newcommand{\Hom}        {\operatorname{\mathsf{Hom}}}   
\newcommand{\ring}[1]    {\mathsf{#1}}                 
\newcommand{\Unit}       {\mathbb{1}}                  
\newcommand{\Aut}        {\operatorname{\mathsf{Aut}}} 
\newcommand{\InnAut}     {\operatorname{\mathsf{InnAut}}} 
\newcommand{\Der}        {\operatorname{\mathsf{Der}}} 
\newcommand{\I}          {\mathrm{i}}
\newcommand{\acts}       {\mathbin{\triangleright}}
\newcommand{\sweedler}[1] {{\scriptscriptstyle{(#1)}}}
\newcommand{\opp}         {\mathrm{opp}}
\newcommand{\twist}[1]   {\mathsf{#1}}

\newcommand{\Bimod}[5] {\sideset{^{\scriptscriptstyle{#1}}_{\scriptscriptstyle{#2}}}{^{\scriptscriptstyle{#4}}_{\scriptscriptstyle{#5}}}{\operatorname{#3}}}

\newcommand{\EA}   {\Bimod{}{}{\mathcal{E}}{}{\mathcal{A}}}
\newcommand{\BEA}  {\Bimod{}{\mathcal{B}}{\mathcal{E}}{}{\mathcal{A}}}
\newcommand{\CFB}  {\Bimod{}{\mathcal{C}}{\mathcal{F}}{}{\mathcal{B}}}
\newcommand{\DGC}  {\Bimod{}{\mathcal{D}}{\mathcal{G}}{}{\mathcal{C}}}
\newcommand{\AAA}  {\Bimod{}{\mathcal{A}}{\mathcal{A}}{}{\mathcal{A}}}
\newcommand{\BBPhiA}{\Bimod{}{\mathcal{B}}{\mathcal{B}}{\Phi}{\mathcal{A}}}

\newcommand{\IP}[4]{{\,}_{\scriptscriptstyle{#2}\!\!}\left\langle{{#1}}\right\rangle^{\scriptscriptstyle{#3}}_{\scriptscriptstyle{#4}}}

\newcommand{\SPA}[1]     {\IP{{#1}}{}{}{\mathcal{A}}}
\newcommand{\SPEA}[1]    {\IP{{#1}}{}{\mathcal{E}}{\mathcal{A}}}
\newcommand{\BSP}[1]     {\IP{{#1}}{\mathcal{B}}{}{}}
\newcommand{\ASP}[1]     {\IP{{#1}}{\mathcal{A}}{}{}}
\newcommand{\SPFB}[1]    {\IP{{#1}}{}{\mathcal{F}}{\mathcal{B}}}
\newcommand{\SPFEA}[1]   {\IP{{#1}}{}{\mathcal{F}\otimes\mathcal{E}}{\mathcal{A}}}
\newcommand{\MnASP}[1]   {\IP{{#1}}{M_n(\mathcal{A})}{}{}}

\newcommand{\tensor}[1][{}]{\mathbin{\otimes_{\scriptscriptstyle{#1}}}}

\newcommand{\Pic}      {\operatorname{\mathsf{Pic}}}
\newcommand{\PicH}      {\sideset{}{_{H}}{\operatorname{\mathsf{Pic}}}}
\newcommand{\StrPic}   {\sideset{}{^{\mathrm{str}}}{\operatorname{\mathsf{Pic}}}}
\newcommand{\starPic}  {\sideset{}{^*}{\operatorname{\mathsf{Pic}}}}
\newcommand{\StrPicH}  {\sideset{}{^{\mathrm{str}}_{H}}{\operatorname{\mathsf{Pic}}}}
\newcommand{\starPicH} {\sideset{}{^*_{H}}{\operatorname{\mathsf{Pic}}}}

\theoremheaderfont{\normalfont\bfseries}
\theorembodyfont{\itshape}
\newtheorem{lemma} {Lemma} [section]
\newtheorem{theorem} [lemma] {Theorem}
\newtheorem{definition}[lemma] {Definition}

\theorembodyfont{\rmfamily}
\newtheorem{remark}[lemma]{Remark}

\numberwithin{equation}{section}

\begin{document}

\maketitle

\begin{abstract}
    In this article we discuss some general results on the covariant
    Picard groupoid in the context of differential geometry and
    interpret the problem of lifting Lie algebra actions to line
    bundles in the Picard groupoid approach.
\end{abstract}



%
%

\tableofcontents

%
%

\section{Introduction}
\label{sec:Intro}

In this work we would like to illustrate and exemplify some general
results from \cite{jansen.waldmann:2004a:pre} where the general
framework of a Morita theory which is covariant under a given Hopf
algebra was studied. One main motivation to do so is coming from
(deformation) quantization theory \cite{bayen.et.al:1978a}, see
e.g~\cite{dito.sternheimer:2002a, gutt:2000a} for recent reviews. Here
Morita equivalence provides an important notion of equivalence of
observable algebras \cite{bursztyn.waldmann:2002a, waldmann:2005b}. In
particular, on cotangent bundles the condition for star products to be
Morita equivalent is shown to coincide with Dirac's integrality
condition for magnetic charges of a background magnetic field
\cite{bursztyn.waldmann:2002a} leading to a natural interpretation of
Morita equivalence also in more general situations.

From the differential geometric point of view, it is a natural
question whether all these techniques as developed in
\cite{bursztyn.waldmann:2001a, bursztyn.waldmann:2003a:pre,
  bursztyn.waldmann:2004a, bursztyn.waldmann:2005a:pre,
  waldmann:2005b} can be made compatible with a certain given symmetry
of the underlying manifold. On the purely algebraic level, a fairly
general notion of `symmetry' is that of a Hopf algebra action of a
given Hopf algebra. In \cite{jansen.waldmann:2004a:pre} we studied
this type of symmetry in the general situation.

From this general framework we shall specialize now into two
directions: on one hand, the algebras on which the symmetry acts and
whose Morita theory shall be studied will now be commutative: we are
interested in the algebra of functions $C^\infty(M)$ on a manifold. On
the other hand, the symmetry in question will either be coming from a
Lie group action on $M$ or from a Lie algebra action as its
infinitesimal counterpart.

It is well-known that two commutative algebras are Morita equivalent
if and only if they are isomorphic, see e.g. the textbook
\cite{lam:1999a}, whence for commutative algebras Morita equivalence
seems to be a useless notion. However, this is not true as things
become interesting if one asks in addition in how many ways two
algebras can be Morita equivalent compared to the ways in which they
can be isomorphic. It turns out that in general there are new
possibilities which makes Morita theory interesting even in the
commutative framework.

This phenomenon is precisely encoded in the so-called Picard groupoid
which we shall compute for the case of function algebras. This way, we
find an interesting and non-trivial class of examples illustrating the
general ideas of \cite{jansen.waldmann:2004a:pre}. Moreover, it will
also be of independent interest as we are now able to re-interpret
several well-known problems and results in differential geometry from
a Morita theoretic point of view. Finally, the commutative situation
with the algebras being function algebras $C^\infty(M)$ is expected
to be the starting point for a discussion of Morita equivalence of
star products as in \cite{bursztyn.waldmann:2002a} but now being
compatible with a symmetry of the classical phase space
\cite{jansen.neumaier.waldmann:2005a:pre}.

The article is organized as follows: In
Section~\ref{sec:GeometryAlgebra} we recall some well-known arguments
why one should and how one can pass from a geometric to a more
algebraic description of differential geometry. The next section is
devoted to a general discussion on Morita theory in different
flavours, taking into account specific structures of the algebras in
question. Here we are mainly interested in $^*$-involutions and
notions of positivity. In Section~\ref{sec:Covariant} we add one more
structure to be preserved by Morita theory, namely a symmetry which we
model by a Hopf algebra action. This can be specialized to group
actions and Lie algebra actions. The last section contains some new
material, namely the explicit computation of a certain part of the Lie
algebra covariant Picard group.

\medskip

\noindent
\textbf{Acknowledgments:} It is a pleasure for me to thank the
organizers and in particular Michel Cahen and Willy Sarlet for their
kind invitation to the 20th International Workshop on Differential
Geometric Methods in Theoretical Mechanics in Ghent where the content
of this work was presented. Moreover, I would like to thank Stefan
Jansen and Nikolai Neumaier for valuable discussions and comments on
the manuscript.

%
%

\section{From Geometry to Algebra}
\label{sec:GeometryAlgebra}

In some sense, differential geometric methods in mathematical physics
correspond mainly to classical theories: Hamiltonian mechanics on a
symplectic or Poisson manifold $M$ is one prominent example. On the
other hand, quantum theories require a more algebraic approach: here
the uncertainty relations in physics are modelled mathematically by
non-trivial commutation relations between observables in some
noncommutative algebra, the observable algebra. Thus
\emph{quantization} in a very broad sense can be understood as the
passage from geometric to noncommutative algebraic structures. An
intermediate step is of course to encode the geometric structures on
$M$ in algebraic terms based on the commutative algebra of functions
$C^\infty(M)$ where, in view of applications to quantization, it is
convenient to consider complex-valued functions.

Then it is a folklore statement (Milnor's exercise) that one can
recover the smooth manifold $M$ from the $^*$-algebra $C^\infty(M)$.
More specifically: every $^*$-homomorphism $\Phi: C^\infty(M)
\longrightarrow C^\infty(N)$ between function algebras is actually of
the form $\Phi = \phi^*$ with some smooth map $\phi: N \longrightarrow
M$ between the underlying manifolds, see
e.g.~\cite{grabowski:2003a:pre, mrcun:2003a:pre} for a recent
discussion. Thus the category of $^*$-algebras with $^*$-homomorphisms
as morphisms becomes relevant to differential geometry.

The `dictionary' to translate geometric to algebraic terms, which is
also one of the cornerstones of Connes' noncommutative geometry
\cite{connes:1994a}, can be extended in various directions. We mention
just one further example also relevant to Morita theory: by the
well-known Serre-Swan theorem, see e.g.\cite{swan:1962a}, the
(complex) vector bundles $E \longrightarrow M$ correspond to finitely
generated projective modules over the algebra $C^\infty(M)$ via $E
\leftrightarrow \Gamma^\infty(E)$. In more geometric terms this means
that for any vector bundle there exists another vector bundle $F
\longrightarrow M$ such that $E \oplus F$ is a trivial vector bundle.
This is of course the key to relate the algebraic $K_0$-theory of
$C^\infty(M)$ to the topological $K^0$-theory of $M$.

Let us now turn to Morita theory. Its first motivation came from the
question what one can say about two algebras $\mathcal{A}$ and
$\mathcal{B}$ provided one knows that their categories of (left)
modules are equivalent, see \cite{morita:1958a, lam:1999a}. Clearly,
in view of applications to quantum mechanics a good understanding of
the more specific modules given by $^*$-representations of the
observable algebras on (pre) Hilbert spaces is crucial for any
physical interpretation. As we shall not start to define what a
reasonable category of modules over the $^*$-algebra $C^\infty(M)$
should be ---though this can perfectly be done, see
e.g.~\cite{bursztyn.waldmann:2003a:pre, schmuedgen:1990a,
  waldmann:2005b} and references therein--- we take a different
motivation which will lead essentially to the same structures. The
idea is to take the category of $^*$-algebras, keep the objects and
enhance the notion of morphisms.  This can be expected to be
interesting as for function algebras we already know what the
`ordinary' morphisms are: pull-backs by smooth maps. Thus a
generalization would lead to a generalization of smooth maps between
manifolds, when we translate things back using our `dictionary'. In
particular, it might happen that algebras become isomorphic in this
new, enhanced category (which will turn out to be not the case for
function algebras) and one might have more `automorphisms' of a given
algebra (which will indeed be the case for function algebras).  In
general, the invertible morphisms in a category form a (large)
groupoid in the obvious sense which is called the Picard groupoid of
the category.  Thus a major step in understanding the whole category
is to consider its Picard groupoid of invertible arrows first.

In principle, the whole idea should be familiar from geometric
mechanics as one example of enhancing a category by allowing more
general morphisms is given by the symplectic `category': first one
considers symplectic manifolds as objects and symplectomorphisms as
morphisms. Though this is a reasonable choice to look at, it turns out
to be rather boring as the choice for the morphisms is too
restrictive.  More interesting is the `category' where one considers
morphisms to be canonical relations, see
e.g.\cite{bates.weinstein:1995a}. However, this is no longer an honest
category since the composition of morphisms is only defined when
certain technical requirements like clean intesections of the
canonical relations are fulfilled. Nevertheless this `symplectic
category' is by far more interesting now.

Other examples are the Morita theory for (integrable) Poisson
manifolds by Xu \cite{xu:1991a} as well as the Morita theory of Lie
groupoids, see e.g.~\cite{moerdijk.mrcun:2003a}.

%
%

\section{Morita Equivalence in Different Flavours}
\label{sec:MoritaFlavours}

After having outlined the general ideas in the previous section we
should start being more concrete now. As warming up we discuss the
`enhancing of the category' for the category of unital algebras with
usual algebra morphisms first, see e.g.~\cite{benabou:1967a,
  lam:1999a} for this classical approach.

Here the generalized morphisms are the \emph{bimodules}: For two
algebras $\mathcal{A}$ and $\mathcal{B}$ a $(\mathcal{B},
\mathcal{A})$-bimodule $\mathcal{E}$, which we shall frequently denote
by $\BEA$ to indicate that $\mathcal{B}$ acts from the left while
$\mathcal{A}$ acts from the right, is considered as an arrow
$\mathcal{A} \longrightarrow \mathcal{B}$.

Why does this give a reasonable notion of morphisms? In particular, we
have to define the composition of morphisms. Thus let $\BEA$ and
$\CFB$ be bimodules then their tensor product $\CFB
\tensor[\mathcal{B}] \BEA$ over $\mathcal{B}$ is a $(\mathcal{C},
\mathcal{A})$-bimodule and hence an arrow $\mathcal{A} \longrightarrow
\mathcal{C}$. However, this is not yet an associative composition law
as for three bimodules $\DGC$, $\CFB$, $\BEA$ we have a
\emph{canonical isomorphism}
\begin{equation}
    \label{eq:GCEisGCE}
    \DGC \tensor[\mathcal{C}] 
    \left( \CFB \tensor[\mathcal{B}] \BEA \right)
    \cong
    \left(\DGC \tensor[\mathcal{C}] \CFB \right)
    \tensor[\mathcal{B}] \BEA
\end{equation}
as $(\mathcal{D}, \mathcal{A})$-bimodules but \emph{not equality}. The
way out is to use isomorphism classes of bimodules as arrows instead
of bimodules themselves. Then the tensor product becomes indeed
associative and the isomorphism class of the canonical bimodule $\AAA$
serves as the identity morphism of the object $\mathcal{A}$ since we
use unital algebras for simplicity.

The final restriction we have to impose is that in a category the
morphism space between two objects has to be a set, which is a priori
not clear in our enhanced category. Therefor one should pose
additional constraints on the bimodules like finitely
generatedness. However, we shall ignore these subtleties in the
following as the new notion of isomorphisms in this category will be
unaffected anyway.

However, we still have to show that we really get an extension of our
previous notion of morphisms. Thus let $\Phi: \mathcal{A}
\longrightarrow \mathcal{B}$ be an algebra homomorphism. Then on
$\mathcal{B}$ we define a right $\mathcal{A}$-module structure by $b
\cdot_{\Phi} a = b \Phi(a)$ and obtain a bimodule $\BBPhiA$. Its
isomorphism class is denoted by $\ell(\Phi)$. It is easy to see that
$\ell(\Phi \circ \Psi) = \ell(\Phi) \circ \ell(\Psi)$ and
$\ell(\id_{\mathcal{A}})$ is the class of $\AAA$ whence our previous
notion of morphisms is indeed contained in the new one.

If we denote this new category by $\mathsf{ALG}$ then two unital
algebras $\mathcal{A}$ and $\mathcal{B}$ are called \emph{Morita
  equivalent} iff they are isomorphic in $\mathsf{ALG}$. Without going
into the details this is equivalent to the existence of a certain
bimodule which is `invertible' with respect to the composition
$\tensor$. In fact, such bimodules can be characterized rather
explicitly, see e.g.~\cite{lam:1999a}.

The isomorphism classes of these invertible bimodules constitute now
the \emph{Picard groupoid} of this category $\mathsf{ALG}$ which we
shall denote by $\Pic$. The invertible arrows from $\mathcal{A}$ to
$\mathcal{B}$ are denoted by $\Pic(\mathcal{B}, \mathcal{A})$ while
the isotropy group of this groupoid at the local unit $\mathcal{A}$ is
denoted by $\Pic(\mathcal{A})$, the \emph{Picard group} of
$\mathcal{A}$.

The map $\ell$ induces now a group homomorphism such that
\begin{equation}
    \label{eq:ExactInnAutAutPic}
    1 
    \longrightarrow \InnAut(\mathcal{A}) 
    \longrightarrow \Aut(\mathcal{A})
    \stackrel{\ell}{\longrightarrow} \Pic(\mathcal{A})
\end{equation}
is exact, whence in the commutative case, the automorphism group of
$\mathcal{A}$ is a subgroup of the Picard group $\Pic(\mathcal{A})$.
Finally, it can be shown that for commutative $\mathcal{A}$, the exact
sequence \eqref{eq:ExactInnAutAutPic} is split whence
\begin{equation}
    \label{eq:Split}
    \Pic(\mathcal{A}) = 
    \Aut(\mathcal{A}) \ltimes \Pic_{\mathcal{A}}(\mathcal{A}),
\end{equation}
where the subgroup $\Pic_{\mathcal{A}}(\mathcal{A})$ consists of the
symmetric invertible bimodules, i.e. those where $a \cdot x = x \cdot
a$ for all $x \in \mathcal{E}$ and $a \in \mathcal{A}$. Then
$\Pic_{\mathcal{A}}(\mathcal{A})$ is called the \emph{commutative} or
\emph{static Picard group}, see e.g.~\cite{bursztyn.waldmann:2004a,
  bursztyn.weinstein:2004a} for a discussion and further references.

It is a well-known theorem in Morita theory that for unital algebras
$\mathcal{A}$, $\mathcal{B}$ the equivalence bimodules $\BEA$ are
certain finitely generated projective right $\mathcal{A}$-modules such
that $\Hom_{\mathcal{A}}(\EA) \cong \mathcal{B}$.  Coming back to our
example $\mathcal{A} = C^\infty(M)$ we see, using the Serre-Swan
theorem, that the only candidates for the symmetric self-equivalence
bimodules are the sections $\Gamma^\infty(L)$ of a complex \emph{line
  bundle}. In fact, it turns out that $\Gamma^\infty(L)$ is indeed
invertible with inverse given by the class of $\Gamma^\infty(L^*)$
since $\Gamma^\infty(L) \tensor[C^\infty(M)] \Gamma^\infty(L) \cong
\Gamma^\infty(L^* \otimes L) \cong C^\infty(M)$ as
$C^\infty(M)$-bimodules. This shows that the static Picard group of
$C^\infty(M)$ is just the `geometric' Picard group, i.e. the group of
isomorphism classes of complex line bundles with the tensor product as
multiplication. Using the Chern class to classify complex line bundles
then gives according to \eqref{eq:Split}
\begin{equation}
    \label{eq:PicCinftyM}
    \Pic(C^\infty(M)) = \mathrm{Diffeo}(M) \ltimes
    \check{\mathrm{H}}^2(M, \mathbb{Z}),
\end{equation}
where the semidirect product structure comes from the usual action of
diffeomorphisms on $\check{\mathrm{H}}^2(M, \mathbb{Z})$.  In general,
all Morita equivalence bimodules $\BEA$ for $\mathcal{A} =
C^\infty(M)$ are isomorphic to some $\Gamma^\infty(E)$ with a vector
bundle $E \longrightarrow M$ of non-zero fibre dimension. Moreover,
$\mathcal{B}$ has to be isomorphic to
$\Gamma^\infty(\mathsf{End}(E))$. Thus for function algebras
$C^\infty(M)$ we have a complete description of the Picard groupoid.

We shall now specialize our notion of Morita equivalence: we have
already argued that the $^*$-involution of $C^\infty(M)$ should be
taken into account when having applications to quantization in mind.
Moreover, one can include notions of positivity into Morita theory.
One defines a linear functional $\omega: \mathcal{A} \longrightarrow
\mathbb{C}$ to be \emph{positive} if $\omega(a^*a) \ge 0$ for all $a
\in \mathcal{A}$. Then an element $a \in \mathcal{A}$ is called
\emph{positive} if $\omega(a) \ge 0$ for all positive linear
functionals $\omega$ of $\mathcal{A}$, see \cite{waldmann:2004a,
  waldmann:2005b, bursztyn.waldmann:2003a:pre, schmuedgen:1990a} for a
detailed discussion. It is clear that for applications to quantum
theories such notions of positive functionals are crucial as they
encode expectation value functionals and hence the physical
\emph{states} for the observable algebra.

In particular, for $\mathcal{A} = C^\infty(M)$ one finds that positive
linear functionals are precisely the integrations with respect to
compactly supported positive Borel measures. This follows essentially
from Riesz' representation theorem, see
\cite[App.~B]{bursztyn.waldmann:2001a}. From this it immediately
follows that $f \in C^\infty(M)$ is positive iff $f(x) \ge 0$ for all
$x \in M$, whence the above, purely algebraic definition reproduces
the usual notion.

We can now state the definition of $^*$-Morita equivalence
\cite{ara:1999a} and strong Morita equivalence bimodules, see
\cite{rieffel:1974b} as well as \cite{landsman:1998a} for Rieffel's
original formulation in the context of $C^*$-algebras and
\cite{bursztyn.waldmann:2003a:pre, bursztyn.waldmann:2001a} for the
general case of $^*$-algebras.  Instead of describing the `enhanced
category' way, we directly give the definition in terms of bimodules
which is entirely equivalent, see \cite{bursztyn.waldmann:2003a:pre}.
\begin{definition}
    \label{definition:MoritaBimodules}
    A $^*$-Morita equivalence bimodule $\BEA$ is a $(\mathcal{B},
    \mathcal{A})$-bimodule together with inner products
    \begin{equation}
        \label{eq:Ainner}
        \SPA{\cdot,\cdot}: \mathcal{E} \times \mathcal{E}
        \longrightarrow \mathcal{A}
    \end{equation}
    and
    \begin{equation}
        \label{eq:Binner}
        \BSP{\cdot,\cdot}: \mathcal{E} \times \mathcal{E}
        \longrightarrow \mathcal{B}
    \end{equation}
    such that for all $x, y, z \in \mathcal{E}$, $a \in \mathcal{A}$
    and $b \in \mathcal{B}$ we have:
    \begin{enumerate}
    \item $\SPA{\cdot,\cdot}$ (resp. $\BSP{\cdot,  \cdot}$) is linear
        in the right (resp. left) argument.
    \item $\SPA{x, y \cdot a} = \SPA{x, y} a$ and $\BSP{b \cdot x, y}
        = b \BSP{x, y}$.
    \item $\SPA{x, y} = {\SPA{y, x}}^*$ and $\BSP{x, y} = {\BSP{y, x}}^*$.
    \item $\SPA{\cdot, \cdot}$ and $\BSP{\cdot, \cdot}$ are
        non-degenerate.
    \item $\SPA{\cdot, \cdot}$ and $\BSP{\cdot, \cdot}$ are full.
    \item $\SPA{x, b \cdot y} = \SPA{b^* \cdot x, y}$ and $\BSP{x, y
          \cdot a} = \BSP{x \cdot a^*, y}$.
    \item $\BSP{x, y} \cdot z = x \cdot \SPA{y, z}$.
    \end{enumerate}
    If in addition the inner products are completely positive then
    $\BEA$ is called a strong Morita equivalence bimodule.
\end{definition}
Here $\SPA{\cdot, \cdot}$ is called \emph{full} if the
$\mathbb{C}$-span of all elements $\SPA{x, y}$ is the whole algebra
$\mathcal{A}$; in general these elements constitute a $^*$-ideal.
Moreover, $\SPA{\cdot, \cdot}$ is called \emph{completely positive} if
for all $n \in \mathbb{N}$ and all $x_1, \ldots, x_n \in \mathcal{E}$
the matrix $(\SPA{x_i, x_j}) \in M_n(\mathcal{A})$ is positive in the
$^*$-algebra $M_n(\mathcal{A})$.

The composition of bimodules is again the tensor product where on
$\CFB \tensor[\mathcal{B}] \BEA$ the $\mathcal{A}$-valued inner
product is now defined by Rieffel's formula
\begin{equation}
    \label{eq:Rieffel}
    \SPFEA{x \otimes \phi, y \otimes \psi}
    =
    \SPEA{\phi, \SPFB{x, y} \cdot \psi},
\end{equation}
and analogously for the $\mathcal{C}$-valued inner product. It is then
a non-trivial theorem that this is indeed completely positive again,
if the inner products on $\mathcal{E}$ and $\mathcal{F}$ have been
completely positive \cite{bursztyn.waldmann:2003a:pre}. Passing to
isometric isomorphism classes one can show that this gives a groupoid:
the $^*$-Picard groupoid $\starPic$ and the strong Picard groupoid
$\StrPic$, respectively. In particular, the local unit at
$\mathcal{A}$ is given by the isometric isomorphism class of $\AAA$
equipped with the inner products
\begin{equation}
    \label{eq:CanonicalInnerProd}
    \SPA{a, b} = a^*b
    \quad
    \textrm{and}
    \quad
    \ASP{a, b} = ab^*.
\end{equation}
More generally, $\mathcal{A}^n$, viewed as $(M_n(\mathcal{A}),
\mathcal{A}$)-bimodule equipped with the canonical inner products
\begin{equation}
    \label{eq:SPAcan}
    \MnASP{x, y} = \sum_{i=1}^n x \cdot \SPA{y, \cdot} 
    \quad
    \textrm{and}
    \quad
    \SPA{x, y} = \sum_{i=1}^n x_i^* y_i
\end{equation}
implements the strong Morita equivalence between $\mathcal{A}$ and
$M_n(\mathcal{A})$.

Since we simply can forget the additional structures we obtain
canonical groupoid morphisms
\begin{equation}
    \label{eq:Triangle}
    \bfig
    \Vtriangle<500,300>[\StrPic`\starPic`\Pic;``]
    \efig,
\end{equation}
which have been studied in \cite{bursztyn.waldmann:2003a:pre}: in
general, none of them is surjective nor injective, even on the level
of the Picard groups.

The geometric interpretation of the inner products is that they
correspond to \emph{Hermitian fiber metrics} on the corresponding line
bundles or vector bundles, respectively: Indeed, this can be seen
easily from the very definitions.  Since up to isometry there is only
one positive Hermitian fiber metric on a given line bundle we have in
the case of $\mathcal{A} = C^\infty(M)$
\begin{equation}
    \label{eq:StrPicCinftyM}
    \StrPic(C^\infty(M)) 
    = \mathrm{Diffeo}(M) \ltimes \check{H}^2(M, \mathbb{Z}) 
    = \Pic(C^\infty(M)).
\end{equation}

\begin{remark}
    \label{remark:WeirdRing}
    In the approach of \cite{bursztyn.waldmann:2003a:pre,
      bursztyn.waldmann:2001a, bursztyn.waldmann:2004a,
      bursztyn.waldmann:2005a:pre} one main point was to replace the
    real numbers $\mathbb{R}$ by an arbitrary \emph{ordered ring}
    $\ring{R}$ and $\mathbb{C}$ by the ring extension $\ring{C} =
    \ring{R}(\I)$ with $\I^2 = -1$. This allows to include also the
    formal star product algebras from deformation quantization into
    the game. They are defined as algebras over the formal power
    series $\mathbb{C}[[\lambda]]$.  Surprisingly, essentially all of
    the constructions involving positivity go through without
    problems.
\end{remark}

%
%

\section{The Covariant Situation}
\label{sec:Covariant}

Let us now pass to the covariant situation: we want to incorporate
some given symmetry of the $^*$-algebras in question. Here we have two
main motivations and examples from differential geometry: First, a
smooth action $\Phi: M \times G \longrightarrow M$ of a Lie group $G$
on $M$, where by convention we choose a right action in order to have
a left action $g \mapsto \Phi_g^*$ on $C^\infty(M)$ by
$^*$-automorphisms. Second, as infinitesimal version of $\Phi$, a Lie
algebra action, i.e. a Lie algebra homomorphism $\varphi: \mathfrak{g}
\longrightarrow \mathfrak{X}(M)$ from a real finite dimensional Lie
algebra $\mathfrak{g}$ into the Lie algebra of real vector fields,
which correspond to the $^*$-derivations of $C^\infty(M)$.

In order to formalize and unify both situations it is advantageous to
consider Hopf $^*$-algebras and their actions on algebras. Thus let
$H$ be a \emph{Hopf $^*$-algebra}, i.e. a unital $^*$-algebra with a
coassociative coproduct $\Delta$, a counit $\epsilon$ and an antipode
$S$ such that $\Delta: H \longrightarrow H \otimes H$ as well as
$\epsilon: H \longrightarrow \mathbb{C}$ are $^*$-homomorphisms and
$S(S(g^*)^*) = g$ for all $g \in H$, see
e.g.~\cite[Sect.~IV.8]{kassel:1995a}. For the coproduct we shall use
Sweedler's notation $\Delta(g) = g_\sweedler{1} \otimes
g_\sweedler{2}$.

The two geometric examples we want to discuss are now encoded in the
following Hopf $^*$-algebras:

First, recall that any group $G$ defines its group algebra
$\mathbb{C}[G]$ which becomes a Hopf $^*$-algebra by setting
$\Delta(g) = g \otimes g$, $\epsilon(g) = 1$ and $S(g) = g^{-1} = g^*$
for $g \in G \subseteq \mathbb{C}[G]$. In this case $H =
\mathbb{C}[G]$ is even \emph{cocommutative}, i.e. $\Delta =
\Delta^{\opp}$ where the opposite coproduct is defined by
$\Delta^{\opp}(g) = g_\sweedler{2} \otimes g_\sweedler{1}$.

Second, for any real Lie algebra the complexified universal enveloping
algebra $U_\mathbb{R}(\mathfrak{g}) \tensor[\mathbb{R}] \mathbb{C} =
U_\mathbb{C}(\mathfrak{g})$ becomes a Hopf $^*$-algebra by setting
$\Delta(\xi) = \xi \otimes \Unit + \Unit \otimes \xi$, $\epsilon(\xi)
= 0$ and $S(\xi) = - \xi = \xi^*$ together with the resulting
extensions to all of $U_\mathbb{C}(\mathfrak{g})$. Again,
$U_{\mathbb{C}}(\mathfrak{g})$ is cocommutative.

The situation that a group $G$ acts by $^*$-automorphisms on a
$^*$-algebra as well as a Lie algebra representation by
$^*$-derivations can be unified in terms of Hopf $^*$-algebras as
follows: A \emph{$^*$-action} $\acts$ of $H$ on $\mathcal{A}$ is a
bilinear map $\acts: H \times \mathcal{A} \longrightarrow \mathcal{A}$
such that $g \acts (h \acts a) = (gh) \acts a$ and $\Unit_H \acts a =
a$, i.e.  $\mathcal{A}$ is a left $H$-module, and $g \acts (ab) =
(g_\sweedler{1} \acts a)(g_\sweedler{2} \acts b)$, $g \acts
\Unit_{\mathcal{A}} = \epsilon(g) \Unit_{\mathcal{A}}$, and $(g \acts
a)^* = S(g)^* \acts a^*$ for all $g, h \in H$ and $a, b \in
\mathcal{A}$.  Then it is well-known and easy to see that for our two
examples $\mathbb{C}[G]$ and $U_{\mathbb{C}}(\mathfrak{g})$ this
indeed generalizes and unifies the action by $^*$-automorphisms and
$^*$-derivations, respectively. The interesting relations are all
encoded in the different coproducts.

In principle and probably even more naturally, one should consider
coactions instead of actions of $H$, see
e.g.~\cite[Sect.~III.6]{kassel:1995a}. Nevertheless, we stick to the
more intuitive point of view where $H$ `acts'.

Now suppose we have $^*$-algebras $\mathcal{A}$, $\mathcal{B}$ with a
$^*$-action of $H$. Let furthermore $\BEA$ be a strong or $^*$-Morita
equivalence bimodule. Then we call the bimodule \emph{$H$-covariant}
if there is an $H$-module structure on $\mathcal{E}$ denoted by
$\acts$, too, such that we have the following compatibilities
\begin{equation}
    \label{eq:aactsbx}
    g \acts (b \cdot x) 
    = (g_\sweedler{1} \acts b) \cdot (g_\sweedler{2} \acts x)
\end{equation}
\begin{equation}
    \label{eq:gactsxa}
    g \acts (x \cdot a) 
    = (g_\sweedler{1} \acts x) \cdot (g_\sweedler{2} \acts a)
\end{equation}
\begin{equation}
    \label{eq:gactsBSP}
    g \acts \BSP{x, y} =
    \BSP{g_\sweedler{1} \acts x, S(g_\sweedler{2})^* \acts y}
\end{equation}
\begin{equation}
    \label{eq:gactsSPA}
    g \acts \SPA{x, y} =
    \SPA{S(g_\sweedler{1})^* \acts x, g_\sweedler{2} \acts y}
\end{equation}
for all $x, y \in \mathcal{E}$, $a \in \mathcal{A}$, $b \in
\mathcal{B}$ and $g, h \in H$. Of course, in the case of
ring-theoretic Morita theory one only requires \eqref{eq:aactsbx} and
\eqref{eq:gactsxa}. Taking isometric isomorphism classes also
respecting the action of $H$ gives the $H$-covariant flavours of the
Picard groupoids, denoted by $\PicH$, $\starPicH$, and $\StrPicH$,
respectively. Since we can successively forget the additional
structures we get the following commuting diagram of canonical
groupoid morphisms:
\begin{equation}
    \label{eq:BigDiagram}
    \bfig
    \Vtriangle(0,350)<500,200>[\StrPicH`\starPicH`\Pic_H;``]
    \Vtriangle(0,0)/@{>}|\hole`>`>/<500,200>[\StrPic`\starPic,`\Pic;``]
    \morphism(0,525)<0,-250>[`;]
    \morphism(500,325)<0,-250>[`;]
    \morphism(1000,525)<0,-250>[`;]
    \efig
\end{equation}

Now let us interpret the diagram on the level of Picard groups and in
our geometric situation: Let e.g. $H = U_{\mathbb{C}}(\mathfrak{g})$
and $\mathcal{A} = C^\infty(M)$ be as before, equipped with an action
of $\mathfrak{g}$ by $^*$-derivations. Then the kernel and the image
of the group morphism
\begin{equation}
    \label{eq:PicHtoPic}
    \PicH(\mathcal{A}) \longrightarrow \Pic(\mathcal{A})
\end{equation}
encodes on which line bundles we can \emph{lift} the
$\mathfrak{g}$-action and if so, in how many different ways up to
isomorphism.  Analogously, in the strong situation one requires in
addition compatibility with the Hermitian fiber metric. The case of $H
= \mathbb{C}[G]$ leads to the question of existence and uniqueness of
liftings of the group action on $M$ to a group action on $L$ by vector
bundle automorphisms. In the strong case one requires the lift in
addition to be unitary with respect to the fiber metric. All this can
easily be seen from the compatibility requirements \eqref{eq:aactsbx},
\eqref{eq:gactsxa}, \eqref{eq:gactsBSP}, and \eqref{eq:gactsSPA}
applied to our situation.

Clearly, all these lifting problems are very natural questions in
differential geometry and have been discussed by various authors, see
in particular \cite{palais.stewart:1961a, stewart:1961a,
  kostant:1970a, lashof:1991a}, whence one can rely on the techniques
developed there. Even though our approach does not give essential new
techniques to attack the (in general quite difficult) lifting problem,
it shines some new light on it and unreveals some additional structure
of the problem, namely the groupoid structures together with the
canonical groupoid morphisms~\eqref{eq:BigDiagram}. Moreover, this
point of view embeds the lifting problem in some larger and completely
algebraic context since neither the $^*$-algebras have to be
commutative nor has the Hopf $^*$-algebra to be cocommutative.  As
remarked already, we can even replace $\mathbb{R}$ and $\mathbb{C}$ by
$\ring{R}$ and $\ring{C}$, respectively, and incorporate in particular
the formal star product algebras from deformation quantization as
well.

%
%

\section{The case of a Lie algebra}
\label{sec:LieAlgebra}

In this last section we consider the case of $H =
U_{\mathbb{C}}(\mathfrak{g})$ more closely and develop some general
results from \cite{jansen.waldmann:2004a:pre} slightly further.

Assume that $\BEA$ is a $^*$-Morita equivalence bimodule which allows
for a lift of the actions of $H$ on $\mathcal{A}$ and $\mathcal{B}$.
Then it was shown in \cite[Thm~4.14]{jansen.waldmann:2004a:pre} in
full generality that the possible lifts are parametrized by the
following group $U(H, \mathcal{A})$:

We consider linear maps $\Hom(H, \mathcal{A})$ with the usual
convolution product given by $(\twist{a} * \twist{b})(g) =
\twist{a}(g_\sweedler{1})\twist{b}(g_\sweedler{2})$. This makes
$\Hom(H, \mathcal{A})$ an associative algebra with unit $\twist{e}(g)
= \epsilon(g) \Unit_{\mathcal{A}}$, see
e.g.~\cite[Sect.~III.3]{kassel:1995a}. Then we consider the following
conditions for $\twist{a} \in \Hom(H, \mathcal{A})$
\begin{equation}
    \label{eq:normalization}
    \twist{a}(\Unit_{H}) = \Unit_{\mathcal{A}},
\end{equation}
\begin{equation}
    \label{eq:cocylce}
    \twist{a}(gh) 
    = \twist{a}(g_\sweedler{1}) (g_\sweedler{2} \acts \twist{a}(h))
    \quad
    \textrm{for all}
    \quad
    g, h \in H,
\end{equation}
\begin{equation}
    \label{eq:central}
    (g_\sweedler{1} \acts b) \twist{a}(g_\sweedler{2})
    =
    \twist{a}(g_\sweedler{1}) (g_\sweedler{2} \acts b)
    \quad
    \textrm{for all}
    \quad
    b \in \mathcal{A}, g \in H,
\end{equation}
\begin{equation}
    \label{eq:unitary}
    \twist{a}(g_\sweedler{1})
    \twist{a}\left(S(g_\sweedler{2}^*)\right)^* 
    = \epsilon(g) \Unit_{\mathcal{A}}
    \quad
    \textrm{for all}
    \quad
    g \in H.
\end{equation}
Then $U(H, \mathcal{A})$ is defined to be the subset of those
$\twist{a} \in \Hom(H, \mathcal{A})$ which satisfy
\eqref{eq:normalization}--\eqref{eq:unitary} and it turns out that
$U(H, \mathcal{A})$ is a group with respect to the convolution product
\cite[App.~A]{jansen.waldmann:2004a:pre}. Moreover, for unitary
central elements $c \in U(\mathcal{Z}(\mathcal{A}))$ one defines
$\hat{c} \in U(H, \mathcal{A})$ by $\hat{c}(g) = c (g \acts c^{-1})$.
Then one obtains the exact sequence
\begin{equation}
    \label{eq:exactsequence}
    1 
    \longrightarrow U(\mathcal{Z}(\mathcal{A}))^H 
    \longrightarrow U(\mathcal{Z}(\mathcal{A})) 
    \stackrel{\widehat{\;\;}}{\longrightarrow}
    U(H, \mathcal{A})
\end{equation}
and the image $\widehat{U(\mathcal{Z}(\mathcal{A}))} \subseteq U(H,
\mathcal{A})$ is a central and hence normal subgroup. Thus we can
define the group $U_0(H, \mathcal{A}) = U(H, \mathcal{A}) \big/
\widehat{U(\mathcal{Z}(\mathcal{A}))}$.

The parametrization of all possible lifts is obtained by a free and
transitive group action $\acts \mapsto \acts^{\twist{b}}$ of $U(H,
\mathcal{B})$ on the set of lifts given by
\begin{equation}
    \label{eq:actsb}
    g \acts^{\twist{b}} x = 
    \twist{b}(g_\sweedler{1}) \cdot (g_\sweedler{2} \acts x),
\end{equation}
where $\twist{b} \in U(H, \mathcal{B})$, $g \in H$ and $x \in
\mathcal{E}$. In fact, for $H$-covariantly $^*$-Morita equivalent
algebras $\mathcal{A}$ and $\mathcal{B}$ we have $U(H, \mathcal{A})
\cong U(H, \mathcal{B})$. Moreover, $\acts^{\twist{b}}$ and $\acts$
give isomorphic actions iff $\twist{b} = \hat{c}$ for some $c \in
U(\mathcal{Z}(\mathcal{B}))$ whence the isomorphism classes of lifts
are parametrized by the group $U_0(H, \mathcal{B}) \cong U_0(H,
\mathcal{A})$ which acts freely and transitively via \eqref{eq:actsb}
on the isomorphism classes of lifts.

While the above characterization works in full generality we want to
specialize now to Lie algebra actions where $H =
U_{\mathbb{C}}(\mathfrak{g})$. First, it follows from
\cite[Prop.~A.7]{jansen.waldmann:2004a:pre} that
$U(U_{\mathbb{C}}(\mathfrak{g}), \mathcal{A}) =
U(U_{\mathbb{C}}(\mathfrak{g}), \mathcal{Z}(\mathcal{A}))$ and hence
$U_0(U_{\mathbb{C}}(\mathfrak{g}), \mathcal{A}) =
U_0(U_{\mathbb{C}}(\mathfrak{g}), \mathcal{Z}(\mathcal{A}))$ since
$U_{\mathbb{C}}(\mathfrak{g})$ is cocommutative. Thus the values of
$\twist{a} \in U(U_{\mathbb{C}}(\mathfrak{g}), \mathcal{A})$ are
automatically central, essentially by \eqref{eq:central}. Moreover,
the groups $U(U_{\mathbb{C}}(\mathfrak{g}), \mathcal{A})$ and
$U_0(U_{\mathbb{C}}(\mathfrak{g}), \mathcal{A})$ are abelian. Since
the center $\mathcal{Z}(\mathcal{A})$ is invariant under the
$\mathfrak{g}$-action (by derivations!) we can restrict $\twist{a} \in
U(U_{\mathbb{C}}(\mathfrak{g}), \mathcal{A})$ to $\mathfrak{g}
\subseteq U_{\mathbb{C}}(\mathfrak{g})$ and obtain a
Chevalley-Eilenberg cochain $\alpha = \twist{a}\big|_{\mathfrak{g}}
\in C^1_{\mathrm{CE}}(\mathfrak{g}, \mathcal{Z}(\mathcal{A}))$.
Evaluating the conditions \eqref{eq:cocylce}, \eqref{eq:central} and
\eqref{eq:unitary} on elements $\xi, \eta \in \mathfrak{g}$ we find by
a simple computation the following lemma:
\begin{lemma}
    \label{lemma:cocycle}
    The restriction gives an injective group homomorphism
    \begin{equation}
        \label{eq:Restrict}
        U(U_{\mathbb{C}}(\mathfrak{g}), \mathcal{A}) \ni \twist{a}
        \mapsto
        \alpha = \twist{a}\big|_{\mathfrak{g}}
        \in
        Z^1_{\mathrm{CE}}\left(
            \mathfrak{g},
            \mathcal{Z}(\mathcal{A})_{\mathrm{anti Hermitian}}
        \right)
    \end{equation}
    into the Chevalley-Eilenberg one-cocycles with values in the anti
    Hermitian central elements of $\mathcal{A}$.
\end{lemma}
The injectivity easily follows from successively applying
\eqref{eq:cocylce}.

Conversely, given $\alpha \in Z^1_{\mathrm{CE}}(\mathfrak{g},
\mathcal{Z}(\mathcal{A})_{\mathrm{anti Hermitian}})$ we can construct
an element $\twist{a} \in U(U_{\mathbb{C}}(\mathfrak{g}),
\mathcal{A})$ with $\twist{a}\big|_{\mathfrak{g}} = \alpha$: Let
$T^k_{\mathbb{C}}(\mathfrak{g})$ denote the $k$-th complexified tensor
power of $\mathfrak{g}$ and define $\twist{a}^{(k)}:
T^k_{\mathbb{C}}(\mathfrak{g}) \longrightarrow \mathcal{A}$
inductively by
\begin{equation}
    \label{eq:Twistafromalpha}
    \twist{a}^{(0)} = \Unit_{\mathcal{A}}
    \quad
    \textrm{and}
    \quad
    \twist{a}^{(k)}(\xi \otimes Y)
    =
    \alpha(\xi) \twist{a}^{(k-1)}(Y) + \xi \acts \twist{a}^{(k-1)}(Y)
    \quad
    \textrm{for}
    \quad
    k \ge 1,
\end{equation}
where $Y \in T^{k-1}_{\mathbb{C}}(\mathfrak{g})$. Then a lenghty but
straightforward computation using $\delta_{\mathrm{CE}} \alpha = 0$
shows that $a = \sum_{k=0}^\infty a^{(k)}$ passes to the universal
enveloping algebra $U_{\mathbb{C}}(\mathfrak{g})$, viewed as a
quotient of $T_{\mathbb{C}}^\bullet(\mathfrak{g})$ in the usual way,
and fulfills \eqref{eq:normalization} to \eqref{eq:unitary}. Thus we
have:
\begin{theorem}
    \label{theorem:Iso}
    The map \eqref{eq:Restrict} is an isomorphism of abelian groups.
\end{theorem}

Note that this is in some sense surprising as the condition for
$\alpha$ to be a cocycle is linear while the condition
\eqref{eq:cocylce} for $\twist{a}$ is highly \emph{non-linear}. It
only becomes linear when evaluated on $\xi, \eta \in \mathfrak{g}
\subseteq U_{\mathbb{C}}(\mathfrak{g})$ thanks to the fact that these
elements are primitive, i.e. satisfy $\Delta(\xi) = \xi \otimes \Unit
+ \Unit \otimes \xi$. Thus a simplification like in
Theorem~\ref{theorem:Iso} cannot be expected for more non-trivial Hopf
$^*$-algebras.

Moreover, under the identification \eqref{eq:Restrict} the elements
$\hat{c}$ give just the cocycles $\hat{c}(\xi) = c (\xi \acts c^{-1})$
as usual. Note that in general $\widehat{U(\mathcal{Z}(\mathcal{A}))}
\subseteq Z^1_{\mathrm{CE}}(\mathfrak{g},
\mathcal{Z}(\mathcal{A})_{\mathrm{anti Hermitian}})$ are \emph{not}
CE-coboundaries. Thus, if we want to relate
$U_0(U_{\mathbb{C}}(\mathfrak{g}), \mathcal{A})$ to Lie algebra
cohomology we have to assume an additional structure for
$\mathcal{A}$:
\begin{definition}
    \label{definition:Exp}
    Let $\mathcal{A}$ be a unital $^*$-algebra. Then an
    exponential function $\exp$ is a map $\exp:
    \mathcal{Z}(\mathcal{A}) \longrightarrow \mathcal{Z}(\mathcal{A})$
    such that
    \begin{enumerate}
    \item $\exp(a+b) = \exp(a)\exp(b)$,
    \item $\exp(0) = \Unit_{\mathcal{A}}$,
    \item $D \exp(a) = \exp(a) Da$,
    \item $\exp(a^*) = \exp(a)^*$,
    \end{enumerate}
    for all $a, b \in \mathcal{Z}(\mathcal{A})$ and $D \in
    \Der(\mathcal{A})$.
\end{definition}
Note that $D \in \Der(\mathcal{A})$ induces an outer derivation
$D\big|_{\mathcal{Z}(\mathcal{A})}$ of the center.

We shall now assume that $\mathcal{A}$ has an exponential function
where our motivating example is of course $\mathcal{A} = C^\infty(M)$
with the usual exponential.

The first trivial observation is that for $a \in
\mathcal{Z}(\mathcal{A})$ we have
\begin{equation}
    \label{eq:hatexpdelta}
    \widehat{\exp(a)}(\xi) = - (\delta_{CE} a)(\xi),
\end{equation}
and for $a = - a^* \in \mathcal{Z}(\mathcal{A})_{\mathrm{anti
    Hermitian}}$ we clearly have $\exp(a) \in
U(\mathcal{Z}(\mathcal{A}))$. Thus in this case
$\widehat{U(\mathcal{Z}(\mathcal{A}))}$ contains all anti Hermitian
CE-coboundaries in $Z^1_{\mathrm{CE}}(\mathfrak{g},
\mathcal{Z}(\mathcal{A})_{\mathrm{anti Hermitian}})$. Note however,
that in general, $\widehat{U(\mathcal{Z}(\mathcal{A}))}$ is strictly
larger. To measure this we consider those elements in
$U(\mathcal{Z}(\mathcal{A}))$ which are not in the image of $\exp$: we
define the abelian group
\begin{equation}
    \label{eq:HdRInteger}
    \mathrm{H}^1_{\mathrm{dR}}
    (\mathcal{Z}(\mathcal{A}), 2 \pi \I \mathbb{Z})
    =
    \frac{U(\mathcal{Z}(\mathcal{A}))}
    {\exp\left(\mathcal{Z}(\mathcal{A})_{\mathrm{anti
              Hermitian}}\right)},
\end{equation}
where the left hand side is of course only a symbol. However, for
$\mathcal{A} = C^\infty(M)$ we obtain indeed the $2\pi\I$-integral
first de Rham cohomology of $M$ by the right hand side of
\eqref{eq:HdRInteger} which motivates our notation. Since
$U(\mathcal{Z}(\mathcal{A}))$ is mapped via $c \mapsto \hat{c}$ into
the cocycles $Z^1_{\mathrm{CE}}(\mathfrak{g},
\mathcal{Z}(\mathcal{A})_{\mathrm{anti Hermitian}})$ and
$\exp(\mathcal{Z}(\mathcal{A})_{\mathrm{anti Hermitian}})$ gives the
coboundaries via \eqref{eq:hatexpdelta} we obtain a well-defined
induced map
\begin{equation}
    \label{eq:hatInduced}
    \mathrm{H}^1_{\mathrm{dR}}
    (\mathcal{Z}(\mathcal{A}), 2 \pi \I \mathbb{Z})
    \stackrel{\widehat{\;\;}}{\longrightarrow}
    H^1_{\mathrm{CE}}\left(
        \mathfrak{g},
        \mathcal{Z}(\mathcal{A})_{\mathrm{anti Hermitian}}
    \right).
\end{equation}
Collecting all the results we obtain the following statement:
\begin{theorem}
    \label{theorem:Unull}
    Assume $\mathcal{A}$ has an exponential function. Then the
    restriction \eqref{eq:Restrict} induces a canonical group
    isomorphism
    \begin{equation}
        \label{eq:Unull}
        U_0(U_{\mathbb{C}}(\mathfrak{g}), \mathcal{A})
        \cong
        \frac{H^1_{\mathrm{CE}}\left(
              \mathfrak{g},
              \mathcal{Z}(\mathcal{A})_{\mathrm{anti Hermitian}}
          \right)}
        {\widehat{\mathrm{H}^1_{\mathrm{dR}}
            (\mathcal{Z}(\mathcal{A}), 2 \pi \I \mathbb{Z})}}.
\end{equation}
\end{theorem}

In particular, this applies to $\mathcal{A} = C^\infty(M)$ whence we
obtain the full classification of the inequivalent lifts of the Lie
algebra action to line bundles. Note that in general,
$U_0(U_{\mathbb{C}}(\mathfrak{g}), \mathcal{A})$ does not depend on
the line bundle itself but is universal for all line bundles.

%
%

\begin{footnotesize}

\begin{thebibliography}{10}

\bibitem {ara:1999a}
{\sc Ara, P.: }\newblock {\em {M}orita equivalence for rings with involution}.
\newblock Alg. Rep. Theo.  {\bf 2} (1999), 227--247.

\bibitem {bates.weinstein:1995a}
{\sc Bates, S., Weinstein, A.: }\newblock {\em Lectures on the Geometry of
  Quantization}.
\newblock Berkeley Mathematics Lecture Notes 8, Berkeley, 1995.

\bibitem {bayen.et.al:1978a}
{\sc Bayen, F., Flato, M., Fr{{\o}}nsdal, C., Lichnerowicz, A., Sternheimer,
  D.: }\newblock {\em Deformation Theory and Quantization}.
\newblock Ann. Phys.  {\bf 111} (1978), 61--151.

\bibitem {benabou:1967a}
{\sc B{\'e}nabou, J.: }\newblock {\em Introduction to Bicategories}.
\newblock In: {\em Reports of the Midwest Category Seminar},   1--77.
  Springer-Verlag, 1967.

\bibitem {bursztyn.waldmann:2001a}
{\sc Bursztyn, H., Waldmann, S.: }\newblock {\em Algebraic Rieffel Induction,
  Formal Morita Equivalence and Applications to Deformation Quantization}.
\newblock J. Geom. Phys.  {\bf 37} (2001), 307--364.

\bibitem {bursztyn.waldmann:2002a}
{\sc Bursztyn, H., Waldmann, S.: }\newblock {\em The characteristic classes of
  {M}orita equivalent star products on symplectic manifolds}.
\newblock Commun. Math. Phys.  {\bf 228} (2002), 103--121.

\bibitem {bursztyn.waldmann:2003a:pre}
{\sc Bursztyn, H., Waldmann, S.: }\newblock {\em Completely positive inner
  products and strong {M}orita equivalence}.
\newblock Preprint (FR-THEP 2003/12)  {\bf math.QA/0309402} (September 2003),
  36 pages.
\newblock To appear in Pacific J. Math.

\bibitem {bursztyn.waldmann:2004a}
{\sc Bursztyn, H., Waldmann, S.: }\newblock {\em Bimodule deformations,
  {P}icard groups and contravariant connections}.
\newblock K-Theory  {\bf 31} (2004), 1--37.

\bibitem {bursztyn.waldmann:2005a:pre}
{\sc Bursztyn, H., Waldmann, S.: }\newblock {\em Induction of Representations
  in Deformation Quantization}.
\newblock Preprint  {\bf math.QA/0504235} (April 2005), 13 pages.
\newblock Contribution to the conference proceedings for the Keio Workshop
  2004.

\bibitem {bursztyn.weinstein:2004a}
{\sc Bursztyn, H., Weinstein, A.: }\newblock {\em Picard groups in {P}oisson
  geometry}.
\newblock Moscow Math. J.  {\bf 4} (2004), 39--66.

\bibitem {connes:1994a}
{\sc Connes, A.: }\newblock {\em Noncommutative Geometry}.
\newblock Academic Press, San Diego, New York, London, 1994.

\bibitem {dito.sternheimer:2002a}
{\sc Dito, G., Sternheimer, D.: }\newblock {\em Deformation quantization:
  genesis, developments and metamorphoses}.
\newblock In: {\sc Halbout, G. (eds.): }\newblock {\em Deformation
  quantization}, vol.~1 in {\em IRMA Lectures in Mathematics and Theoretical
  Physics},   9--54. Walter de Gruyter, Berlin, New York, 2002.

\bibitem {grabowski:2003a:pre}
{\sc Grabowski, J.: }\newblock {\em Isomorphisms of algebas of smooth functions
  revisited}.
\newblock Preprint  {\bf math.DG/0310295} (2003), 5 pages.

\bibitem {gutt:2000a}
{\sc Gutt, S.: }\newblock {\em Variations on deformation quantization}.
\newblock In: {\sc Dito, G., Sternheimer, D. (eds.): }\newblock {\em
  Conf{\'e}rence Mosh{\'e} Flato 1999. Quantization, Deformations, and
  Symmetries}, {\em Mathematical Physics Studies} no. {\bf 21},   217--254.
  Kluwer Academic Publishers, Dordrecht, Boston, London, 2000.

\bibitem {jansen.neumaier.waldmann:2005a:pre}
{\sc Jansen, S., Neumaier, N., Waldmann, S.: }\newblock {\em Covariant {M}orita
  Equivalence of Star Products}.
\newblock In preparation.

\bibitem {jansen.waldmann:2004a:pre}
{\sc Jansen, S., Waldmann, S.: }\newblock {\em The {$H$}-covariant strong
  Picard groupoid}.
\newblock Preprint (FR-THEP 2004/16)  {\bf math.QA/0409130 v2} (2004), 50
  pages.
\newblock To appear in J. Pure Appl. Alg.

\bibitem {kassel:1995a}
{\sc Kassel, C.: }\newblock {\em Quantum Groups}.
\newblock {\em Graduate Texts in Mathematics} no. {\bf 155}.
\newblock Springer-Verlag, New York, Berlin, Heidelberg, 1995.

\bibitem {kostant:1970a}
{\sc Kostant, B.: }\newblock {\em Quantization and Unitary Representation. Part
  I: Prequantization}.
\newblock In: {\sc Taam, C.~T. (eds.): }\newblock {\em Lectures in Modern
  Analysis and Application}, vol. 170 in {\em Lecture Notes in Mathematics},
  87--208. Springer-Verlag, Berlin, 1970.

\bibitem {lam:1999a}
{\sc Lam, T.~Y.: }\newblock {\em Lectures on Modules and Rings}, vol. 189 in
  {\em Graduate Texts in Mathematics}.
\newblock Springer-Verlag, Berlin, Heidelberg, New York, 1999.

\bibitem {landsman:1998a}
{\sc Landsman, N.~P.: }\newblock {\em Mathematical Topics between Classical and
  Quantum Mechanics}.
\newblock {\em Springer Monographs in Mathematics}.
\newblock Springer-Verlag, Berlin, Heidelberg, New York, 1998.

\bibitem {lashof:1991a}
{\sc Lashof, R.: }\newblock {\em Equivariant Prequantization}.
\newblock In: {\sc Dazord, P., Weinstein, A. (eds.): }\newblock {\em Symplectic
  Geometry, Groupoids, and Integrable Systems}, vol.~20 in {\em Mathematical
  Sciences Research Institute Publications},   193--207. Springer-Verlag, New
  York, Berlin, Heidelberg, 1991.

\bibitem {moerdijk.mrcun:2003a}
{\sc Moerdijk, I., Mr{\v{c}}un, J.: }\newblock {\em Introduction to Foliations
  and {L}ie Groupoids}.
\newblock {\em Cambridge studies in advanced mathematics} no. {\bf 91}.
\newblock Cambridge University Press, Cambridge, UK, 2003.

\bibitem {morita:1958a}
{\sc Morita, K.: }\newblock {\em Duality for modules and its applications to
  the theory of rings with minimum condition}.
\newblock Sci. Rep. Tokyo Kyoiku Daigaku Sect. A  {\bf 6} (1958), 83--142.

\bibitem {mrcun:2003a:pre}
{\sc Mr{\v{c}}un, J.: }\newblock {\em On Isomorphisms of Algebras of Smooth
  Functions}.
\newblock Preprint  {\bf math.DG/0309179} (2003), 4 pages.

\bibitem {palais.stewart:1961a}
{\sc Palais, R.~S., Stewart, T.~E.: }\newblock {\em The Cohomology of
  Differentiable Transformation Groups}.
\newblock Amer. J. Math.  {\bf 83} (1961), 623--644.

\bibitem {rieffel:1974b}
{\sc Rieffel, M.~A.: }\newblock {\em Morita equivalence for {$C^*$}-algebras
  and {$W^*$}-algebras}.
\newblock J. Pure. Appl. Math.  {\bf 5} (1974), 51--96.

\bibitem {schmuedgen:1990a}
{\sc Schm{\"{u}}dgen, K.: }\newblock {\em Unbounded Operator Algebras and
  Representation Theory}, vol.~37 in {\em Operator Theory: Advances and
  Applications}.
\newblock Birkh{\"{a}}user Verlag, Basel, Boston, Berlin, 1990.

\bibitem {stewart:1961a}
{\sc Stewart, T.~E.: }\newblock {\em Lifting Group Actions in Fibre Bundles}.
\newblock Ann. Math.  {\bf 74} (1961), 192--198.

\bibitem {swan:1962a}
{\sc Swan, R.~G.: }\newblock {\em Vector bundles and projective modules}.
\newblock Trans. Amer. Math. Soc.  {\bf 105} (1962), 264--277.

\bibitem {waldmann:2004a}
{\sc Waldmann, S.: }\newblock {\em The Picard Groupoid in Deformation
  Quantization}.
\newblock Lett. Math. Phys.  {\bf 69} (2004), 223--235.

\bibitem {waldmann:2005b}
{\sc Waldmann, S.: }\newblock {\em States and Representation Theory in
  Deformation Quantization}.
\newblock Rev. Math. Phys.  {\bf 17} (2005), 15--75.

\bibitem {xu:1991a}
{\sc Xu, P.: }\newblock {\em Morita Equivalence of Poisson Manifolds}.
\newblock Commun. Math. Phys.  {\bf 142} (1991), 493--509.

\end{thebibliography}

\end{footnotesize}

\end{document}